\documentclass[
]{ceurart}

\usepackage{listings}
\begin{document}

\copyrightyear{2026}
\copyrightclause{Copyright for this paper by its authors.
 Use permitted under Creative Commons Attribution License 4.0 
 International (CC BY 4.0).}

\conference{C\&ESAR 2026 (33rd Computer \& Electronics Security Application Rendez-vous)}

\title{Building Resilience in Immature Organisations Against Cyber Crises Through Immersive Training}



\author[1]{Marc Parenthoën}[%
orcid=0000-0001-7432-7119,
email=marc.parenthoen@univ-poitiers.fr,
]
\cormark[1]

\author[2,3]{José Manuel Castillo}[%
orcid=0000-0002-5110-3486,
email=jose.manuel.castillo.pimentel@univ-poitiers.fr,
]

\author[2]{Nicolas Louveton}[%
orcid=0000-0002-9021-2589,
email=nicolas.louveton@univ-poitiers.fr,
]

\author[4]{Alexandre Wets}[%
email=alexandre.wets@diateam.net
  ]

\author[4]{Corentin Tuot}[%
email=corentin.tuot@diateam.net
  ]
  
\author[5]{Hadrien Bourgogne}[%
email=hadrien.bourgogne@crisalyde.fr
  ]

\author[5]{Clément Judek}[%
email=clement.judek@crisalyde.fr
  ]

\author[5]{Selim Miled}[%
email=selim.miled@crisalyde.fr
  ]

\address[1]{XLim CNRS UMR 7252, Université de Poitiers}
\address[2]{CeRCA CNRS UMR 7295, Université de Poitiers}
\address[3]{Perseus UR 7312, Université de Lorraine}
\address[4]{DIATEAM, \url{https://www.diateam.net/}, Brest}
\address[5]{Crisalyde, \url{https://crisalyde.fr/}, Pantin}

\cortext[1]{Corresponding author.}


\begin{abstract}
The RéSISTeCC exercise protocol is the result of three years of interdisciplinary work spanning incident response, crisis management, and cognitive ergonomics. The RéSISTeCC project stands for resilience through strategic and technical immersive simulation of cyber crises and was funded by NextGenerationEU and France 2030 through the DEFFINUM call for proposals. The latest version of this protocol offers a seven-hour cyber crisis training session designed for organisations with limited cybersecurity experience. Its purpose is to assess and test crisis management processes from both technical and strategic perspectives. The exercise protocol involves a strategic unit and a technical unit responsible for cyber crisis management, which are required to collaborate and coordinate their actions simultaneously. Only the technical unit is immersed in a cyber range simulating a generic "LOREM" information system. Both units receive phone calls from the facilitators and can call or visit one another.
 
Traditionally, cyber crisis management exercises require careful consideration during their design due to their socio-technical complexity. Likewise, their management and execution typically depend on a predefined and fixed schedule of events developed in advance. In the case of the RéSISTeCC exercise protocol it has been possible to establish a dynamic exercise management approach that adapts pedagogical objectives and related injects to the participants’ actual responses and actions throughout the exercise.

This paper provides an assessment of the activities carried out during the testing of the RéSISTeCC protocol, describing the needs in cyber range topologies for IT services and how these infrastructures should be populated with data according to the type of target organisations: local governments, small and medium-sized enterprises (SMEs), healthcare and long-term care facilities, fire and rescue services. We delve into the conception and progression of the exercise from its initial stages right through to crisis management and the importance of these different phases for situation awareness, once the attack occurs. We explain real-time participants’ performance measurements and how they are used to contribute to the facilitators’ pool of injects based on pedagogical objectives and on the actual state of the IT infrastructure in relation to the cyberattack and the actions of those involved.

These lessons learnt are based on the design, implementation, and testing of 28 exercises involving crisis response teams ranging in size from 2 to 30 people.

\end{abstract}

\maketitle

\section{Introduction}
Crisis management involves a complex set of actions whose main objective is to address the uncertainty and disruption affecting a given structure. As Babin \cite{Babin2020} indicates, this set of actions is carried out by a unit called the crisis unit which serves as the central hub for coordinating actions to manage the situation and involves diverse actors, objectives, functions, and informational needs \cite{Rogova2009}. The notion of Performance in crisis management depends on how effectively an organisation operates in terms of detecting the signals of a forthcoming crisis, preparing for, and preventing it, recognising and containing it, and recovering and learning post-crisis  \cite{Tokakis2019}. In this perspective, if organisations can be proactive, adaptable and agile, they can better manage crises \cite{Raut2022}. Both public and private companies are increasingly interested in training their employees to manage such crises.

The aim of this paper is to share how we have developed cyber crisis exercises within the framework of the RéSISTeCC project, whose objectives are to establish an immersive platform / environment enabling participant collaboration, develop a set of exercises tailored to local authorities and small socio-economic entities, and consolidate a continuing education catalogue in cyber crisis management \cite{Castillo2024}

At the core of the RéSISTeCC project lies the use of a cyber range (CR) as an immersive learning environment. In general terms, a CR consists of a set of virtual machines designed to replicate an information system within an isolated and controlled environment. Such a simulation enables participants to assess and develop their cybersecurity competencies in a safe and controlled setting, without exposing operational or production systems to risk.

The purpose of using an immersive environment is to enable participants to experience a cyber crisis and test their strategic and technical response mechanisms across a range of scenarios. This immersive training is structured around a predefined scenario comprising a sequence of simulated events designed and monitored by the pedagogical team. During the exercises, the scenarios primarily focused on attacks aimed at compromising critical information or disrupting essential services provided by the participating organisations.

This paper presents a synthesis of the collective reflection and lessons learnt from the design and implementation of the RéSISTeCC project. More specifically, it aims to consolidate and assess the knowledge acquired regarding the design and delivery of cyber crisis exercises supported by immersive cyber range technologies and specifically tailored to non-expert audiences.

A distinctive feature of the RéSISTeCC approach is that, unlike approaches focusing exclusively on technical cybersecurity experts or, conversely, on non-technical business stakeholders, it brings together technical and strategic teams within the same immersive environment, combining a realistic simulation of the IT system with a realistic simulation of the organisational impacts that arise from the incident. The objective is not only to strengthen individual competencies, but also to foster collective capabilities in coordination, communication, and decision-making during a cyber crisis. This interdisciplinary approach results from the collaboration of specialists in technical incident response training, experts in the strategic management of cyber crises, and researchers in psychology and cognitive ergonomics. It is grounded in the design, implementation, and experimentation of 28 cyber crisis exercises involving crisis response teams ranging from 2 to 30 participants.

This study therefore contributes to the broader objective of translating the knowledge throughout the project into the RéSISTeCC exercise protocol, providing methodological guidance for the design and implementation of immersive cyber crisis exercises. Emphasis is placed on the challenges associated with designing exercises for participants with limited cybersecurity expertise, as well as to the interaction and coordination between technical and non-technical stakeholders within the same exercise environment.

This project is particularly relevant given the growing need for practical approaches to designing and delivering cyber crisis exercises that are accessible to non-expert audiences. Despite the increasing use of cyber exercises as a means of strengthening organisational preparedness and resilience, the specific methodological challenges involved in adapting such exercises to non-expert participants, while effectively integrating technical and strategic perspectives, remain insufficiently addressed. By documenting and critically reflecting on the experience gained through RéSISTeCC, this paper seeks to contribute to addressing this gap and to provide transferable knowledge for the development of immersive cyber crisis exercises targeting diverse organisational audiences.

Following a review of the current state of the art in immersive training for cyber crisis management, this article will  proceed to presenting the exercise protocol, which features a scenario  adaptive to  the skills observed in real time. It will then describe the simulated topologies in the cyber range that support the technical aspects of the exercises. And finally, it will set  out the method for dynamically assessing participants’ skills and therefore enabling the scenario to be adapted in real time. A discussion will then take place to summarise the lessons learnt before the conclusion.

\section{State of the art in cyber crisis training}
\label{art}

In cyber crisis management, actions and decisions can be divided into two main areas: the technical incident response, which focuses on managing infrastructure and information systems; and the strategic incident response to the organisational impacts of the cyber incident, which involves the organisation’s core activities, the prioritization of the latter, as well as internal and external communication \cite{AFNOR2022}. 

Currently, various methods are used to develop crisis management or cyber defence skills. These range from traditional approaches such as seminars and workshops to more dynamic approaches such as games and simulation exercises. Angafor et al. \cite{Angafor2020} indicate that the current cyber threat landscape requires a combination of training and educational approaches that allow operators to learn in practical, real-life environments, rather than solely in classrooms. Hence, Immersive learning is one of these methods, using artificial environments to replicate experiences close to reality to develop practical skills. In cybersecurity, one of the most sophisticated immersive learning proposals, or as some call it, hands-on experience, is cyber range training.

Cyber Ranges (CR) \cite{Yamin2020} are computing platforms replicating IT infrastructures: networks, systems, tools, and applications, offering a secure environment for acquiring practical cybersecurity skills \cite{Debatty2019}. Traditionally, training using these virtual machines is aimed at a technical audience with established cybersecurity expertise, especially in cyber incident management. These virtual environments have been used in an international military training context to develop both technical and strategic skills, as well as interactions between the various participants \cite{Smeets2022}; but such bespoke large-scale developments are beyond the needs of small organisations with limited cybersecurity capabilities. However, throughout 2024 and 2025, we have been using these platforms to train entities with low experience in cyber crisis management. 

We are interested in CR as a simulated environment training platform to develop the cyber resilience of such immature entities. As we have already announced in several of our works \cite{Castillo2025,Castillo2026}, CR training for these types of organisations can be an innovative way to understand decision-making, communication between teams, and, above all, to achieve collaboration between technical and non-technical stakeholders during crises.

Training through CR lies in its capacity to elicit a realistic set of behaviours, emotions, communications, and decision-making processes through a fictional scenario, socio-technical interactions, and various events injection \cite{Yamin2020, Paijanen2021}. However, to ensure adequate training using CRs, practitioners must recognise that organising a cyber exercise requires structured planning, design, execution, and evaluation. The available literature shows the existence of various frameworks that seek to evaluate the effectiveness of teams during cybersecurity training. Among them, the NIST CRPT framework \cite{NIST2023} provides a checklist focused on key technical characteristics; the ESCO WG5 framework \cite{ECSO2024}, for its part, focuses more on the simulation environment; and the CRDF framework \cite{Katsantonis2023} proposes a methodology for the industrialisation of learning environments with individualised monitoring.

All these frameworks concentrate on developing technical competencies associated with incident detection and response. However, these skills, although essential in cyber crisis contexts, do not encompass the full range of capabilities required to manage these situations. These frameworks
do not address cyber crisis management from an organisational perspective nor sufficiently consider communication between technical teams and strategic crisis units.

For example, according to the NICE framework, NIST 800-181 \cite{Petersen2020}, cyber crisis education objectives are grouped into one competency area and two primary professional roles. The Cyber Resilience competency area (NF-COM-007) describes the learner’s ability to design, develop, implement, and maintain the reliability of cyber-enabled systems to anticipate, withstand, recover from, and adapt to adverse conditions, attacks, or security compromises. The professional roles associated with these activities are on the one hand, that associated with incident response (PD-WRL-003), responsible for investigating, analysing and responding to cybersecurity incidents in networks, with a total of 106 elements related to the Tasks, Knowledge, Skills (TKS); and  on the other hand, the forensic computing role (PD-WRL-002), responsible for analysing digital evidence derived from incidents to extract useful information that supports the mitigation of vulnerabilities in systems and networks, with a total of 181 TKS. These two roles combined represent more than 400 TKS, which can be intimidating for small organisations, in which no internal team is able to acquire such a high number of competencies. ECSF\footnote{European Cybersecurity Skills Framework (ECSF), working under the guidance of ENISA.\url{https://mta-sts.enisa.europa.eu/ECSF/}} produces the same sort of results. Still, even if the problem is poorly defined, organisations must continue training if they have limited access to TKS acquisition. Since a training session typically lasts seven hours, it is unrealistic to structure scenarios exclusively around TKS; instead, focusing on the roles involved in cyber crisis management is more efficient.

For immature organisations, the Small Business Quick Start Guide \cite{NIST2024} offers a simplified NIST CSF framework, focused on critical tasks from a line-of-business risk perspective. Additionally, and from a technical perspective, a cyber range tailored to small organisations can be classified as part of the hybrid CR family (according to the NIST CRPT guide \cite{NIST2023}), consisting of virtual machines built according to a virtual network infrastructure, with cyber-physical connections via Ethernet devices. In the context of cyber resilience development, the NIST guide points out the need to design ad hoc curricula based on the use case. However, this option is often unfeasible for small organisations due to budget and resource constraints.

According to the ECSO WG5 working group \cite{ECSO2024}, the need for key functionalities in CRs is recognised: internet services, attack execution, user activity, scenario and content development, competency management, data collection and analysis, scoring and reporting, and instructor tools. In the case of a CR tailored to small organisations, rather than replicating a specific organisation, too complex and expensive, the replicable and generic LOREM configuration focuses on the functions, services and situations required to train the expected crisis management and incident response capabilities. Facilitators typically simulate offensive and user activities, while the scenario is predefined, and a competency assessment grid facilitates scoring and reporting.

We could define our initiative as a first iteration of the process of the CRDF framework \cite{Katsantonis2023}, aimed at identifying opportunities for improvement and initiating a process of industrialisation of this immersive learning environment. However, we recognise that this framework does not adequately integrate strategic competencies and interactions between technical and strategic teams, which is essential for a coordinated response to cyber crises.

Last February, ENISA published “The ENISA Cybersecurity Exercise Methodology” \cite{ENISA2026}, which, whilst providing a useful framework for the design and delivery of exercises, nevertheless focuses on the organisation of national or sector-level exercises.

As we were unable to find any clear method in the literature for designing and conducting an immersive cyber crisis management exercise that enhances incident response, business continuity and crisis communication capabilities addressing small organisations with limited cybersecurity capabilities, we developed our own protocol.

\section{The RéSISTeCC protocol}

This section presents the protocol developed for the RéSISTeCC exercises. Its development followed a multidisciplinary and iterative approach, involving engineers, cognitive ergonomics researchers, information systems specialists, and cyber crisis management experts. The protocol was progressively refined throughout the project, based on observations and feedback collected during the field implementation of the exercises. This iterative process enabled the pedagogical team to identify practical challenges and progressively adapt both the exercise design and its underlying methodology.

A key characteristic of the RéSISTeCC exercises is the mobilisation of two complementary units within the participating organisation: a technical unit and a strategic unit. Both units are composed of members of the organisation itself and are expected to interact and coordinate throughout the crisis. This configuration makes it possible to address not only the technical management of a cyber incident, but also the strategic, organisational, and decision-making dimensions of crisis response.

The exercises are structured as narrative scenarios in which facilitators introduce events or stimuli, commonly referred to as injects, that may trigger technical incidents or influence the progression of the crisis. On the one hand, these injects can be structuring, guiding participants towards specific tasks, actions, or behaviours, and on the other, overwhelming, introducing distractions, competing demands, or time pressure to increase uncertainty and cognitive or emotional stress. 

The exercise therefore seeks to reproduce not only the technical dimensions of a cyber crisis, but also its cognitive, emotional, and organisational dimensions. In particular, the exercises aim to reproduce interactions, coordination mechanisms, and decision-making processes that are as close as possible to those encountered in real-world crisis situations. This includes creating conditions under which participants may activate and apply their organisation’s existing internal procedures and response mechanisms.

An important aspect of the RéSISTeCC methodology is the adaptation of each exercise to the participating organisation. Prior to the exercise, each organisation underwent a preliminary interview with the RéSISTeCC pedagogical team. These interviews were used to gather information about the organisation’s structure, activities, information systems, existing crisis-management procedures, and specific cybersecurity challenges. This information was subsequently used to tailor the exercise scenario, the cyber range environment, and the injects to the organisation’s specific context and objectives. Such an adaptation was intended to increase the ecological validity of the exercise by ensuring that the simulated situation was sufficiently consistent with the organisational, technical, and operational reality of the participants.

These ecological needs added to our generic LOREM constraints lead us to distinguish target groups of the exercises.

\subsection{Target groups of the exercises}

Each exercise requires the participating organisation to be placed in a crisis situation. Therefore, a technical and narrative scenario must be developed to outline a disruption affecting at least one of the organisation’s essential activities. These essential activities may vary considerably across different socio-economic systems, while the cost of designing a bespoke exercise may be prohibitive for most entities operating within such systems.

Because of pragmatic constraints and funding procedures, we have restricted the study to entities identified at a regional level prior to the RéSISTeCC project, with the assistance of regional partners involved in cyber risk management, notably the \emph{Agence Nationale de Sécurité des Systèmes d’Information} (ANSSI), the \emph{Agence Regionale de Santé} (ARS) and \emph{Groupements Régionaux d’Appui au Développement de la eSanté} (GRADeS), the CLUSIR Aliénor association, employers’ organisations and a group of Chief Information Security Officers (CISOs) from the business clusters in northern Nouvelle-Aquitaine (France).

We therefore identified four types of organisations, each with its own specific set of exercises, charac­terised by similar core activities: local authorities, SMEs, health and social care establishments (ESMS), and departmental fire and rescue services (in French \emph{Service Départemental d'Incendie et de Secours} - SDIS).

All these organisations have information systems that provide functions relating to human resources, finance and the organisation’s internal processes. The various information systems deployed in the CR are described in section \ref{CR}.

In the end, we carried out 28 exercises lasting between half a day and a full day: this duration was dictated by the organisations with which we were able to conduct the exercises: it was impossible for these small organisations – despite being involved in cyber crisis management training – to devote more time to a crisis exercise requiring the director, business unit managers and IT staff to be fully engaged for the entire duration of the exercise.
\begin{itemize}
\item 3 municipalities with fewer than 5,000 inhabitants (C1, C2, C3),
\item 3 urban areas, including two metropolitan areas (A1, A2, A3), 
\item 2 fire and rescue services (D1, D2), 1 departmental council (D3), 
\item 7 social care establishments (1 hospital group (E1), 3 residential care homes for the elderly (E2, E3, E4), 2 hospitals (E5, E6), 1 specialist care home (E7)) 
\item 12 companies: 2 technopoles (T1, T2), 6 mid-sized enterprises (4 industrial and 2 service-sector: M1, M2, M3, M4, M5, M6), 4 SMEs (IT service providers: P3, P4, P5, P6).
\end{itemize}

We have carried out two main types of exercise protocol: First, a half-day exercise involving a single technical-strategic crisis unit (15 iterations / conducted exercises: C1, C2, C3, E1, E2, E3, P1, P2, P3, P4, P5, M1, M2, M3, M4) and second, a full-day exercise involving two interacting units, one technical and the other strategical (13 iterations / conducted exercises: A1, A2, A3, D1, D2, D3, E4, E5, E6, E7, M5, M6, P6).
Based on our initial half-day trials, we found that half a day is much too short for participants to have time to familiarise themselves sufficiently with the cyber range environment in order to be able to respond effectively to a cyberattack, even for intermediate-level technical teams. Some participants who are not comfortable with computers may even lose motivation. Furthermore, it is impossible to bring together a strategic team for more than half a day.

Having worked with organisations both with and without their own IT department, we ultimately only deal with organisations that have their own IT department or with those whose IT service provider is involved and that openly admits to having failed in preventing a cyberattack. Indeed, the immersive simulation of an information system is only truly useful if IT specialists can take control of the technical infrastructure under a cyber-attack and rehearse the incident response. More generally, they must be able to manage crisis processes for 7 hours; if not, and this is particularly the case for beginners, the cost of the exercise is too high regarding the benefits of immersion \cite{Castillo2026b}.

Now that we have exhibited the target groups of the exercises, we will now look at the way an exercise is conducted, whilst maintaining the technical and strategic coherence of the training programme.

\subsection{Exercise procedure and consistency in strategic and technical coordination}

\begin{figure}[b]
  \caption{The main phases of the RéSISTeCC exercise}
  \label{fig:etapes}
  \centering
  \includegraphics[width=\linewidth-80pt]{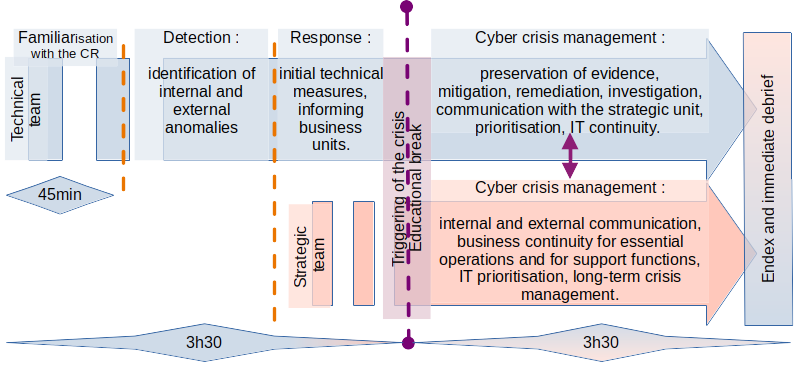}
\end{figure}

The latest version of the protocol exercise lasts approximately seven hours. It begins with the technical team working alone, immersed in the CR. They are gathered in a room, where they are confronted with a technical scenario consisting of a succession of injects that progressively alter the state of the information system, or inform them of its dysfunction. This initial phase is intended to establish a shared understanding of the technical situation, from which organisational consequences may emerge and potentially escalate into a broader crisis (Figure \ref{fig:etapes}). At a later stage, the strategic team is convened, based on information provided by both the technical team and the facilitators to the respective members of the strategic team. Crisis management subsequently relies on structured coordination and information exchange between the two teams.

An introductory phase sets out the objectives, rules and available resources, followed by a 45-minute session on familiarising participants with the information system, (appropriation phase). Preparatory injects are provided to assist with mapping the system (LDAP, Firewall, SIEM, backups, file sharing, application servers, data servers, etc.). In parallel with this initial assessment, and throughout the rest of the exercise -- provided that the network configuration allows it -- the red team facilitator in charge of the cyberattack develops their kill chains, the aim of which is to compromise the essential functions provided by the organisation’s simulated information system. 

Ensuring that the facilitators’ inputs are consistent with the technical reality of the simulated system is no easy task, yet this is an important factor.  On the one hand, consistency is crucial for the sense of \emph{presence} \cite{Zelter92} of the technical participants in the exercise, in the sense that they are immersed in a simulated environment which must enable natural decision-making. On the other hand, it is essential for enhancing the believability of interactions between the technical participants and the strategic participants.

The technical and strategic coherence of the exercise stems from the sharing of information from the CR amongst the exercise facilitators; an incident requiring a technical investigation must correspond to a reality that is simulated. Furthermore, as we do not know in advance what the participants will do when responding to incidents, we have put in place a system to monitor the availability of the essential functionalities characteristic of the type of organisation being simulated. For example, when a business service goes down, a facilitator can trigger a test request from a fictitious user of that service (who is no longer able to work) either to a business manager or to the IT team.

This alignment between business and IT, ensured by the facilitators, enables the interactions between the two groups of participants to be explored. The cognitive effects of an inconsistency are highly disruptive and require the exercise leader to intervene with the technical team. To lower the chance of such impairing situations happening, we rely on an automated support system for facilitators described below.

We may encounter difficulties with certain organisations that are complete novices in crisis manage­ment when it comes to channelling decisions down from the strategic group to the technical group to define IT continuity priorities that support business continuity. When this situation arises, we force an inject as if it had come from the strategic unit, so that the technical unit can carry out this task.

\subsection{Consistency and adaptation of the scenario to the participants’ skills}
\label{adapt}

Unlike traditional exercises, where the timeline is drawn up in advance and mainly unfolds independently from the participants’ actions, our timeline is constructed on the fly based on the participants’ real-time behaviour and performance, with the aim of improving their weakest crisis management skills. These performance metrics are measured using observables based on best practice recommended by experts and linked to the learning objectives for managing cyber-related crises (see section \ref{Percy}).

Based on the observation of participants’ behaviour, performance metrics help to identify which learning objectives still need work and which have been mastered. When a skill appears to be weak following a poor performance, we suggest a series of injects that gradually progress from subtle cues to a clear indication of a problem which needs resolving. If participants appear to be struggling despite the problem being obvious, the exercise is temporarily paused for a brief methodological review (less than 15 minutes) on how to tackle such a problem, after which the exercise continues.

\begin{figure}[h]
  \caption{Automated facilitation support.}
  \label{fig:injects}
  \includegraphics[width=\linewidth-2pt]{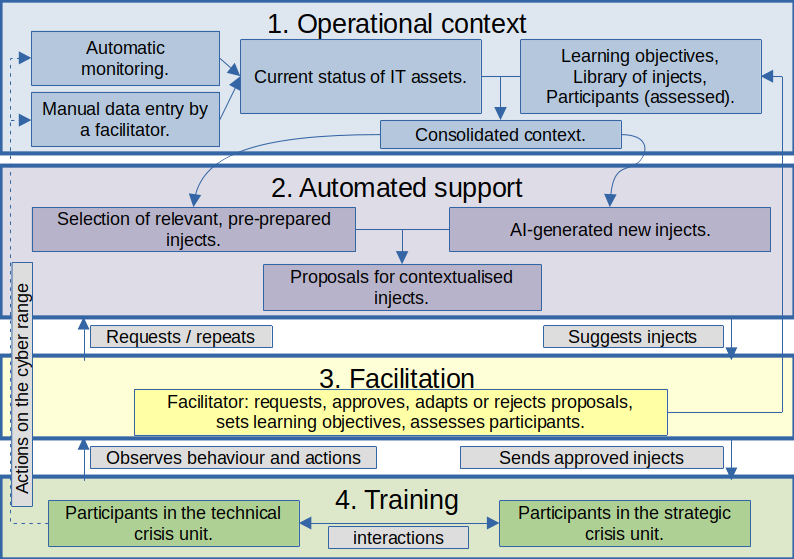}\\[5pt]
  \begin{minipage}{\linewidth-2pt}\em
Based on the status of IT assets (deteriorated or faulty) — updated automatically or manually — and the context of the exercise (1.), automated support selects or uses AI to generate context-specific intervention proposals (2.). The facilitator validates, adapts or rejects these, then adjusts the simulation (3.) according to the reactions observed (4.).
  \end{minipage}
\end{figure}

To help facilitators choose relevant injects, we have developed a system that suggests such potential injects (see Figure \ref{fig:injects}). At present, the support provided to facilitators through the provision of context-specific injects is designed to assist the facilitation team without overriding their judgement. The system developed links the status of technical assets observed during the exercise, the prepared ideal scenario and the players’ reactions to provide, at the right moment, injects that are consistent with the actual state of the information system and the crisis management process.

The status of technical assets can be updated manually or populated automatically by an external monitoring system hosted within the CR. Each change is logged along with its date, source and a comment. Dependencies between technical assets and business activities are also considered: a failure in an IT component can therefore be propagated to the business services that depend on it.

When an asset enters a degraded or unavailable state, the platform sends a real-time event to the monitoring interface. The incident is classified according to its priority — P2 for a degradation, P1 for a failure — and the data feeds already associated with that asset and that state are immediately highlighted.  The facilitator may then request the generation of additional proposals. The generator receives a structured context comprising:
\begin{itemize}
\item the type, domain and context of the exercise;
\item the asset in question, its type and current state;
\item the state transition giving rise to the proposal;
\item the units, functions and participants involved in the exercise.
\end{itemize}

It generates ready-to-use draft alerts: sender, recipient, channel, content, educational purpose, expected responses and trigger time. The tone and level of urgency are tailored to the situation: operational disruption for degraded service, and explicit urgency and business impacts for complete unavailability.

These suggestions are never triggered automatically for the players. They remain drafts that the facilitator can edit, regenerate, discard or publish. Only this last action creates an operational event (inject) within the scenario. This ‘human-in-the-loop’ approach maintains control over the exercise whilst reducing the time needed to devise a believable follow-up.

At the same time, the platform allows participants’ feedback to be recorded, their chat exchanges to be monitored, and call logs, observers’ notes and educational evaluations to be analysed. These indicators provide the facilitator with an insight into the dynamics of the session: decisions made, sticking points, unresolved questions or objectives that were not sufficiently addressed. This allows him to select or adjust the suggested inputs to unlock a crisis unit, increase the pressure, test another communication channel or give the players more time.

The advantage of this architecture is therefore that it combines two complementary loops: 
\begin{itemize}
\item an automatic loop, based on the actual state of the system, 
\item a human loop, based on observation of the players’ behaviour.
\end{itemize}
   
This artefact was used by the facilitators during six exercises (A3, D3, E4, E7, M6 and P6). According to facilitators, it did indeed make facilitation and observation easier. Progress towards the learning objectives remains the responsibility of the instructors, who use the Percy framework described in section \ref{Percy}. 

We have seen how dynamic adaptation is supported by a semi-automatic system. Let us now quickly look at the simulated infrastructure in the cyber range.

\section{Characterisation of generic LOREM information systems}
\label{CR}

The immersive environment for the technical part of the exercise is supported by a CR. We begin by describing the infrastructure and services to be simulated in the CR, and what specific elements are required for the different types of organisations: local authorities, SMEs, ESMS or SDIS.

We draw on the fundamental principle of Chekhov’s gun, derived from cognitive ergonomics: every object and every piece of data present serve a purpose at some point during the exercise \cite{Castillo2026b}. What needs to be included in the CR, and what is sufficient, is therefore solely that which relates to the technical team, namely a manageable infrastructure supporting the organisation’s essential functions as currently operated.

Let us now examine the elements common to all four types of entity and those specific to each.

\subsection{Infrastructure common to all topologies}

\begin{figure}[t]
  \caption{Typical infrastructure diagram}
  \label{fig:carto}
  \centering
  \includegraphics[width=\linewidth-80pt]{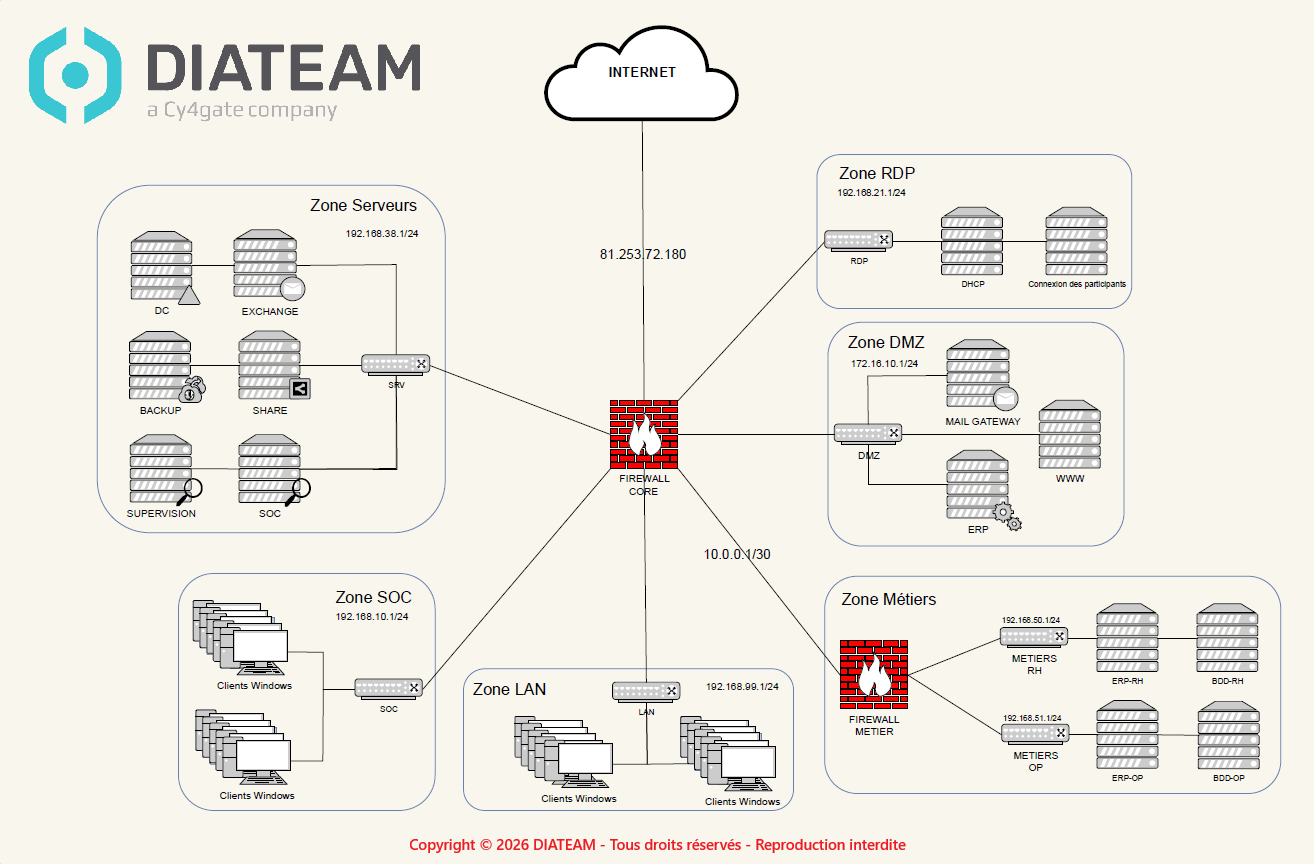}
\end{figure}

The simulated environment is a reference infrastructure (Fig. \ref{fig:carto}) that is fully virtualised within the CR.

This infrastructure is modular and can be adapted to the context of the organisation being trained. It incorporates an Active Directory-type architecture and associated technologies, the organisation’s business services, as well as the main network filtering and security components. Participants interact with it in real time using commonly employed administration protocols, just as they would on a production infrastructure.

When it comes to simulating attackers, tools commonly used by real-world cyber threats are deployed via a pseudo-Internet integrated into the CR. This approach promotes maximum immersion and makes it possible to replicate situations that closely resemble real-world operational conditions.

As part of an exercise, a realistic attack scenario based on credible threats and potential risks is simulated on this virtualised infrastructure. Taking on the roles of staff members or agents within the simulated organisation, participants must detect, analyse and contain the incident by applying the procedures they deem appropriate in line with their usual protocols, to which the facilitators are able to adapt.

For an organization, this technical focus of the exercise is an excellent way to test the effectiveness of its technical and crisis management procedures, while highlighting areas for improvement in the interaction between the operational and strategic plans.

Regarding exercise logistics:
\begin{itemize}
\item When an organisation takes part in an exercise, a dedicated infrastructure is deployed and made available to it.
\item Participants can access this very easily via the DIATEAM Labs web interface.
\item A technical handbook for participants is provided for each exercise.
\item Facilitators can view the status of the simulation environment in real time to adjust the scenario inputs.
\end{itemize}

The data is the same for everyone and consistent between the LDAP and ERP systems. The dates are regenerated for the day of the exercise so that the integrity of the ERP data can be tested in the scenario. For organisations at an intermediate maturity level, traffic is generated from the organisation to various internet services, which makes incident response slightly less straightforward.

We have seen common infrastructure, services and data. Let's talk few words about specific functions.

\subsection{Specific key features}

Whilst human resources and finance are similar from an IT perspective, the activities relating to an organisation’s core purpose are specific to each. We have chosen to assign these specific, essential activities to virtual machines (VMs) with the same name; however, only the connection for administering the VM is operational – no other services or data are hosted by these VMs. In other words, a set of machines are empty shells, each representing a specific essential function of a particular type of organisation; for example: an ICS for an SME, a civil registry or drinking water service for an urban area, or a medical analysis service for a healthcare organisation. These machines are located behind a second firewall.

For the Departmental Fire and Rescue Service (see Figure \ref{fig:carto}), an intervention module and the associated data were developed within the ERP because it was necessary to be able to simulate the IT synchronisation between data from operational fire and rescue functions and that from the support functions for this activity, based on feedback from SDIS that had suffered cyber-attacks.

After these quick descriptions of topologies used in the CR, let’s present the learning objective and performance evaluation.

\section{Performance measures and pedagogical objectives}
\label{Percy}

In this section, we outline the initial version of a guide for assessing performance in relation to the pedagogical objectives that underpin the scenarios in our exercises. 

\begin{table}[ht]
\caption{First list of actions linked to pedagogical objectives (PERCY)}
\label{tab:pedago}
\scriptsize
\begin{tabular}{|p{1.8cm}|p{2.2cm}|p{10cm}|p{1cm}|}
\hline
\textbf{\small Pedagogical objective} & \textbf{\small Intermediate objective} & \textbf{\small Task / Operationalization} & \textbf{\small Crisis Unit} \\
\hline
\multirow{6}{*}{\parbox{1.8cm}{\vspace{1cm}\small Crisis Unit Cyber incident handling}} & 
\multirow{4}{*}{\parbox{2.2cm}{Identification of anomalies and cyber incident origin}} &
Participants identify one or more anomalies or malfunctions in their work structure. & T \\
\cline{3-4}
& & Participants gather additional information regarding these anomalies. & T \\
\cline{3-4}
& & Participants identify the cyber incident(s) that produce the anomalies in their structure. & T \\ & & & \\
\cline{2-4}
& \multirow{5}{*}{\parbox{2.2cm}{Understanding and qualifying the incident}} &
Participants describe and categorize the cyber incident. & T \\
\cline{3-4}
& & Participants determine which activities or workstations are affected by the cyber incident. & T \\
\cline{3-4}
& & Participants identify the types/characteristics of affected data. & T \\
\cline{3-4}
& & The participants (technical, then strategic) make an estimate of the severity based on the issues and threats of the incident. & T/S \\
\cline{3-4}
& & Participants communicate the incident to their technical or IT department and request specific instructions for its management. & T/S \\
\cline{2-4}
& \multirow{2}{*}{\parbox{2.2cm}{Securing the IS}} &
The technical department communicates the security status of the IS and its impact to the structure. & T/S \\
\cline{3-4}
& & Participants follow established procedures or instructions to isolate or protect the compromised system or infrastructure. & T/S \\
\hline
\multirow{5}{*}{\parbox{1.8cm}{\vspace{1cm}\small Constitution and operation of a crisis unit}} &
Activation of the crisis unit & Depending on the severity of the incident (as well as its various indicators), the participants decide to activate the internal crisis management unit. Participants trigger the crisis unit based on indicators. & T/S \\
\cline{2-4}
& \multirow{2}{*}{\parbox{2.2cm}{organisation of the crisis unit}} &
Participants use their internal equipment or equipment available in the room to make decisions (whiteboard, event/action tracking board, logbook, etc.) & T/S \\
\cline{3-4}
& & Participants mobilize their emergency equipment (local telephony, etc.) & S \\
\cline{2-4}
& \multirow{2}{*}{\parbox{2.2cm}{Identification of stakeholders}} &
Participants identify the external organisations that play a role in crisis management & T/S \\
& & Participants notify external and regulatory organisations of the occurrence of a cyber crisis. & T/S \\
\hline
\multirow{9}{*}{\parbox{1.8cm}{\vspace{1cm}\small Decision-making for the strategic response to a cyber-origin crisis.}} &
\multirow{4}{*}{\parbox{2.2cm}{Detection of events within the crisis unit and understanding of the situation}} &
Participants conduct a situation assessment within the crisis cell. & T/S \\ & & & \\
\cline{3-4}
& & Participants formulate hypotheses regarding the direction of the incident, its possible evolution, the reasons for the attack or the attacker's intention, etc. & T/S \\ & & & \\
\cline{2-4}
& Anticipation of future issues & Participants project the impacts of this/these incident(s) in their structure (services, activities, workstations, citizens, etc.) in the short and medium term. & T/S \\
\cline{2-4}
& \multirow{3}{*}{\parbox{2.2cm}{Formulating a collective response to the crisis}} &
Participants define a concrete action plan to continue essential activities in the short term. & T/S \\
\cline{3-4}
& & Participants define a plan to continue supporting essential activities in degraded mode over the medium term. & T/S \\
\cline{3-4}
& & Participants define indicators to return to normal operations and exit the crisis in the long term. & S \\
\cline{2-4}
& \multirow{3}{*}{\parbox{2.2cm}{Operation of the crisis cell}} &
Participants make decisions based on indicators, measurements, and concrete, factual elements. & T/S \\
\cline{3-4}
& & Participants build an artifact to share a common vision. & T/S \\
\cline{3-4}
& & Participants conduct situation updates adapted according to needs. & S \\
\hline
\multirow{6}{*}{\parbox{1.8cm}{\vspace{1mm}\small Management of External Communication and Internal Communication.}} &
\multirow{3}{*}{\parbox{2.2cm}{Internal communication strategy}} &
Participants develop a strategy and plan for internal communication management / strategy &  T/S \\
\cline{3-4}
& & Participants select a relevant information channel appropriate to the situation. &  T/S \\
\cline{3-4}
& & Participants send clear messages aligned with the strategy to the workers in the structure. &  T/S \\
\cline{2-4}
& \multirow{3}{*}{\parbox{2.2cm}{External communication strategy}} &
Participants establish a strategy and plan for managing external communications & S \\
\cline{3-4}
& & Participants check that the message is correctly transmitted to the media and/or external organisations. & S \\
\cline{3-4}
& & Participants adopt a proactive attitude and communication with law enforcement and/or external organisations. & S \\
\hline
\multirow{3}{*}{\parbox{1.8cm}{\vspace{3mm}\small Analysis and remediation of the technical incident}} &
Design / Monitoring of a Business Continuity Plan & Participants implement the IT system continuity procedures, including providing backup equipment, switching to manual (paper-based) processes if necessary, and managing operations via telephone and other resources. & T/S \\
\cline{2-4}
& Analysis and investigation of IoCs & Participants request the preservation of evidence for investigation. & T/S \\
\cline{2-4}
& Design / Monitoring of a Disaster Recovery Plan (DRP) & Participants follow a business recovery plan; if none is available, they begin to develop one based on the information received and the experiences gained during the exercise. & T/S \\
\hline
\multicolumn{4}{|c|}{T: Technical unit ; S: Strategic unit} \\
\hline
\end{tabular}
\end{table}

A list of observable indicators has been drawn up (see Table \ref{tab:pedago}) based on the specific actions expected during a cyber crisis. These indicators are based on the main existing standards for cyber crisis management and are structured within the PERCY (Performance and Response to Cyber Crises) assessment framework. This assessment framework aims to model cognitive and behavioural performance in the context of a cyber crisis simulation.

\subsection{PERCY framework to evaluate cybercrisis immature organisations}

One challenge we noted was that   in this type of training, participants’ actions are highly volatile and difficult to predict, therefore considerable prior work is required to define the specific actions that should be assessed. These limitations are compounded by the difficulties of the existing frameworks when applied in immersive training contexts. For example, as discussed in the state of the art (section \ref{art}), NIST 800-181 \cite{Petersen2020} organizes cyber crisis learning objectives within the area of cyber resilience and into two professional roles: incident response and forensics, which together comprise more than 400 elements of tasks, knowledge, and skills to be developed. In addition to being difficult to operate in a short-term training program, this volume is disproportionate for small immature entities, which lack the human and technical resources necessary to cover these competencies. Similarly, ISO 22301:2019 and ISO/IEC 27005:2022, while offering solid principles for documenting, establishing, and implementing business continuity management or information security system management, are more oriented toward defining policies and processes rather than formulating evaluable learning objectives and/or concrete actions to measure, making them less practical in the context of immersive training. In contrast, AFNOR SPEC 2208 represents a step forward by providing a list of concrete actions to follow in the event of a crippling cyberattack for small immature entities, making it easier to use as a basis for creating learning objectives. However, even this approach lacks a structured framework which would link these actions to clear performance evaluation mechanisms in cyber crisis scenarios. Consequently, there is a need for proposals, such as the Performance and Reaction to a Cyber Crisis Framework, which integrates the operational clarity of standards like AFNOR with a pedagogical and evaluative approach specific to small immature entities in immersive cyber crisis training contexts. Based on the work developed over the past two years, we can present a first list of pedagogical objectives, their definition, and concrete actions that participants must execute during the exercises. Table \ref{tab:pedago} presents the list of pedagogical objectives.
A few of these were examined \citep{Castillo2025}, but we need to improve this list by the expert validation method.

\subsection{Validation method for the PERCY framework}

\begin{figure}[h]
  \caption{Process of validation for PERCY}
  \label{fig:validation}
  \centering
  \includegraphics[width=\linewidth-10pt]{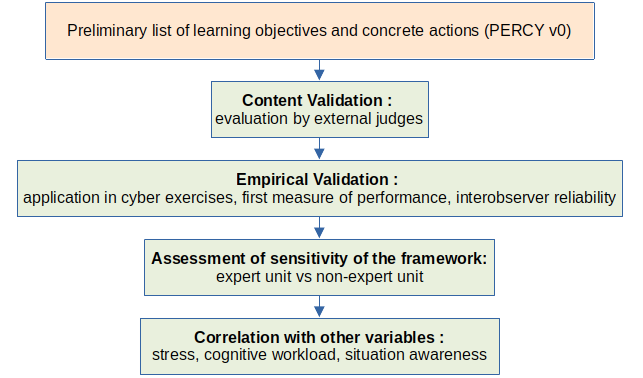}
\end{figure}

Once an initial list of pedagogical objectives has been established, along with the concrete actions that operationalize them, the next step is to validate these objectives. A structured process will be followed to achieve objective validation. Figure \ref{fig:validation} presents the framework validation process.

This validation process comprises several stages. The first stage is validating the framework’s content (validity content), which external judges will carry out. This method consists of presenting, through an online questionnaire, the list of pedagogical objectives and the actions that comprise them to a selected number of professionals and researchers in information systems, cybersecurity, and crisis management. The idea is for the judges to evaluate whether each action described effectively represents the corresponding pedagogical objective. To do this, two aspects will be considered: first, the context of the framework, designed for immersive cyber crisis management training; and second, the time constraints inherent to the training. A Likert-type scale will be used to gather the experts’ opinions, where one indicates that the action does not represent the pedagogical objective and five indicates that the action fully represents it. This will allow the external judges’ assessment to be quantified. Additionally, a space will be provided for judges to leave qualitative comments. This stage will establish a preliminary version of the framework (pedagogical objectives and associated actions). 

Once the preliminary version is established, the second, empirical validation will be carried out. This will consist of assessing whether the framework is relevant for measuring the effectiveness of crisis teams. To do this, different experts will observe a crisis exercise and use specific actions to evaluate the performance of the crisis units. Finally, the evaluators’ perceptions will be collected, and inter-observer reliability will be calculated, thus assessing the consistency of the framework and the measurement instrument. 

The third evaluation will analyse whether the instrument allows differentiation (sensitivity of the instrument) between expert and non-expert groups. It will assess whether the stipulated actions are sufficiently sensitive to distinguish between the two groups. To this end, two crisis units will be compared, each performing the same exercise scenario, ensuring that they are as homogeneous as possible. 

Finally, it will be observed whether the actions, when translated into performance, correlate with other indicators such as stress, mental workload, and situational awareness. Once all these steps are completed, a list of validated actions for each pedagogical objective will be obtained.

\section{General discussion and lessons learnt}

The experience gained through the RéSISTeCC project suggests that the design of immersive cyber crisis exercises cannot be reduced to the technical configuration of a cyber range. Rather, the effectiveness of such training depends on the interaction between the simulated environment, the scenario, the stimuli introduced by facilitators, and the cognitive and organisational characteristics of the participants \cite{Castillo2025, Castillo2026b}.

A critical lesson emerging from the exercises concerns the need to minimise unintended deviations in participants' behaviour. Such deviations may arise from unfamiliarity with the simulated environment (applications included in the cyber range), uncertainty regarding the rules of the exercise, or a mismatch between the scenario and the participants' actual organisational context. While some degree of unpredictability is inherent to crisis situations and can be pedagogically valuable, excessive or irrelevant deviations may interfere with the progression of the exercise and ultimately compromise its learning objectives. The design challenge is therefore not to fully control participants' behaviour, but rather to establish sufficient conditions for their actions to remain within a meaningful and pedagogically relevant space.

Four complementary approaches were adopted in RéSISTeCC to address this issue. First, each exercise was preceded by a preliminary interview with the participating organisation. These interviews allowed the pedagogical team to identify the organisational, technical, and operational characteristics that needed to be represented in the exercise, so that the scenario could represent a dreaded event stemming from a cyber-attack. This information was subsequently used to determine which elements of the information system, internal procedures, organisational roles, and crisis-management practices should be reproduced in the cyber range and incorporated into the scenario. Such preliminary contextualisation was essential to maintain a sufficient degree of fidelity between the simulated environment and the participants' actual working context. This criterion lets us propose 4 types of infrastructure/scenarios answering needs for our whole ecosystem studied.

Second, particular attention was given to the participants' familiarisation with the exercise. A dedicated appropriation phase was introduced before the crisis scenario itself, allowing participants to discover the cyber range environment, familiarise themselves with the available tools and information sources, and understand how the simulated environment was structured. This phase also served to establish the operational rules of the exercise, including communication procedures, interaction between participants, and expected responses to events occurring outside the immediate scope of the simulation. Beyond its technical function, this familiarisation period therefore helped establish a shared understanding of the exercise as a socio-technical environment. It reduced uncertainty unrelated to the learning objectives and allowed participants to devote a greater proportion of their cognitive resources to the crisis itself once the exercise began.

Third, the design of the scenarios progressively incorporated a principle inspired by the narrative concept commonly known as \emph{Chekhov's gun}: elements introduced into a scenario should have a meaningful function in its subsequent development. Applied to cyber crisis exercises, this principle implies that tools, information sources, emails, alerts, databases, and other artefacts presented to participants should contribute either to the progression of the scenario or to its pedagogical objectives. This principle is particularly relevant when attempting to balance realism and mental effort. Adding elements simply to increase the apparent realism of a cyber range may generate irrelevant cognitive demands and encourage participants to explore information that has no meaningful role in the crisis. Conversely, ensuring that introduced elements have a purpose contributes to narrative coherence and helps maintain participants' attention on the relevant decision-making processes \cite{Castillo2025}.

Fourth, given the dynamic nature of crisis scenarios and the potentially unpredictable behaviour of participants, it is necessary to develop tools that support the effective management and adaptation of exercise injects. In the case of RéSISTeCC, an internal tool was tested to assess and adjust the flow of injects according to the progression of the exercise (Section \ref{adapt}). This type of initiative highlights the need for dedicated mechanisms that facilitate the effective execution of scenarios and, in particular, provide support to facilitators in managing and steering the exercise. Such tools can complement the principles of purposeful scenario design by enabling facilitators to dynamically adapt the timing, sequencing, and delivery of injects in response to participants' actions, thereby preserving both narrative coherence and the pedagogical objectives of the exercise.
These observations reinforce the importance of the relationship between the simulated environment and the participants' actual working context. The cyber range should reproduce the relevant characteristics of the organisation's information system with sufficient fidelity to create a meaningful representation of the participants' operational environment, in this way, the main challenge and question is how a generic cyber range can fit with these needs. In this way, the response to  this question does not necessarily imply reproducing the entire information system. Instead, the challenge is to identify and reproduce those technical and organisational elements that are relevant to the learning objectives of the exercise. Similarly, scenarios need to remain believable and sufficiently aligned with the activities, structures, and crisis-management procedures of the participating organisation. 

Excessive abstraction may reduce participants' engagement, whereas unnecessary technical complexity may increase cognitive demands without contributing to the pedagogical objectives. This issue is particularly important when considering the management of injects. They are essential for progressively shaping the evolution of the crisis, but their accumulation could also generate an undesirable level of cognitive saturation. The objective is therefore not to maximise the number or intensity of injects, but to determine an appropriate balance between challenge and cognitive capacity. From this perspective, the design of immersive training can be considered an ergonomic optimisation problem, in which realism, complexity, time pressure, and participants' cognitive resources must be jointly considered. CRs should consequently be regarded not only as technical platforms, but also as socio-cognitive training devices that organise information, tasks, interactions, and contextual cues in ways that are compatible with participants' cognitive and decision-making processes.

The experience accumulated during RéSISTeCC also points to the need to expand the range of injects used to structure the exercises in addition to red-teaming activity. To date, messages, social media posts and telephone calls have been used extensively to represent events and progressively increase the cognitive demands placed on participants. Future iterations could incorporate a broader range of visual injects, including images, videos, screenshots, or other forms of multimedia information. Such injects could provide alternative ways of representing events and potentially increase immersion by reproducing the heterogeneous information sources encountered during real-world cyber crises. However, their introduction should follow the same principle of pedagogical relevance: additional modalities should be incorporated when they support the learning objectives rather than simply increasing the sensory richness of the simulation.

Finally, the RéSISTeCC experience raises a methodological issue concerning the evaluation and generalisation of immersive cyber crisis exercises. The exercises were conducted with organisations characterised by different structures, operational contexts, information systems, and levels of cybersecurity expertise. As a result, the organisational processes and responses observed during the exercises were highly variable. Establishing a fully standardised experimental framework is therefore challenging. In this respect, the RéSISTeCC approach currently lies between a case-study methodology and an experimental approach. This position provides valuable ecological validity, as the exercises are closely connected to real organisational contexts, but it also limits the extent to which results can be directly compared across organisations. 

\section{Conclusion}
To conclude this work, it is worth recalling a thought attributed to Jean Monnet, the French political thinker and statesman: \emph{“People only accept change when they are faced with necessity and only recognize necessity during crises.”}. Applied to cyber crisis management, this statement highlights the importance of creating realistic experiences that reveal how organisations mobilise their competencies and internal processes when confronted with a cyber crisis.

In today’s cybersecurity landscape, no organisation can be considered entirely immune to cyberattacks and their potential consequences. Developing realistic cyber crisis scenarios can therefore help organisations identify weaknesses in their internal processes, coordination mechanisms, and decision-making capabilities, while providing valuable opportunities for organisational learning and preparedness.

We first presented an analysis of the current state of knowledge regarding immersive training in cyber crisis management, showing that the implementation of cyber crisis management exercises in an immersive environment for small, inexperienced organisations remains an unresolved issue.
Then we detailed the RéSISTeCC protocol, which features a scenario that adapts according to the skills observed. After which we described the simulated topologies in the cyber range that support the technical aspects of the exercises; and finally, we set out the method PERCY for dynamically assessing participants’ skills, enabling the scenario to be adapted in real time.
 To this day, the validation of the protocol is only operational on the basis of 6 successful exercises, which were all possible thanks to improving progress achieved over 22 previous exercises.
We will continue to apply cognitive ergonomics in the design and development of exercises to optimise the subtle balance involved in the learning process for cyber crisis management by organisations at the beginner or intermediate stages. To conclude, the RéSISTeCC project was initially conceived to strengthen cyber crisis management capabilities in organisations with limited cybersecurity maturity. However, after a long process of development, experimentation, and refinement, the project has not only contributed to this initial objective but has also generated knowledge and methodological insights that may be valuable to larger and more mature organisations.

\begin{acknowledgments}

The \emph{Résilience par Simulation Immersive Stratégique et Technique de Crise d'origine Cyber} (RéSISTeCC) project, a successful applicant under the DEFFINUM call for proposals, was funded by NextGenerationEU and France 2030. It offers organisations at the beginner to intermediate levels seven-hour cyber crisis exercises to test both technical and strategic crisis management processes. Thanks to the 25 organisations -- local governments, SMEs, health care facilities, social service providers, and fire and rescue services -- which took part in the pilot phase of these exercises between February 2024 and March 2026.

\end{acknowledgments}

\section*{Declaration on Generative AI}
The author(s) have not employed any Generative AI tools.
%

\bibliography{resistecc_caesar}

\begin{thebibliography}{21}
\expandafter\ifx\csname natexlab\endcsname\relax\def\natexlab#1{#1}\fi
\providecommand{\url}[1]{\texttt{#1}}
\providecommand{\href}[2]{#2}
\providecommand{\path}[1]{#1}
\providecommand{\DOIprefix}{doi:}
\providecommand{\ArXivprefix}{arXiv:}
\providecommand{\URLprefix}{URL: }
\providecommand{\Pubmedprefix}{pmid:}
\providecommand{\doi}[1]{\href{http://dx.doi.org/#1}{\path{#1}}}
\providecommand{\Pubmed}[1]{\href{pmid:#1}{\path{#1}}}
\providecommand{\bibinfo}[2]{#2}
\ifx\xfnm\relax \def\xfnm[#1]{\unskip,\space#1}\fi
\bibitem[{Babin et~al.(2020)Babin, Laclemence, Delatour, and
  Richard}]{Babin2020}
\bibinfo{author}{M.~Babin}, \bibinfo{author}{P.~Laclemence},
  \bibinfo{author}{G.~Delatour}, \bibinfo{author}{P.~H. Richard},
\newblock \bibinfo{title}{{La proximit{\'e} cognitive, un facteur d'influence
  de l'activit{\'e} manag{\'e}riale au sein d'une cellule de crise}},
\newblock in: \bibinfo{booktitle}{{Congr{\`e}s Lambda Mu 22 `` Les risques au
  c{\oe}ur des transitions '' (e-congr{\`e}s) - 22e Congr{\`e}s de Ma{\^i}trise
  des Risques et de S{\^u}ret{\'e} de Fonctionnement, Institut pour la
  Ma{\^i}trise des Risques}}, \bibinfo{year}{2020}, pp. \bibinfo{pages}{1--11}.
  \bibinfo{note}{\url{https://hal.science/hal-03454751}}.
\bibitem[{Rogova(2009)}]{Rogova2009}
\bibinfo{author}{G.~Rogova},
\newblock \bibinfo{title}{{Context-awareness} in crisis management},
\newblock in: \bibinfo{booktitle}{Military Communications Conference (MILCOM)},
  \bibinfo{publisher}{IEEE}, \bibinfo{address}{USA}, \bibinfo{year}{2009}, pp.
  \bibinfo{pages}{1--7}.
  \bibinfo{note}{\url{https://doi.org/10.1109/MILCOM.2009.5379827}}.
\bibitem[{Tokakis et~al.(2019)Tokakis, Polychroniou, and
  Boustras}]{Tokakis2019}
\bibinfo{author}{V.~Tokakis}, \bibinfo{author}{M.~Polychroniou},
  \bibinfo{author}{G.~Boustras},
\newblock \bibinfo{title}{Crisis management in public administration: the three
  phases model for safety incidents},
\newblock \bibinfo{journal}{Safety Science} \bibinfo{volume}{113}
  (\bibinfo{year}{2019}) \bibinfo{pages}{37--43}.
  \bibinfo{note}{\url{https://doi.org/10.1016/j.ssci.2018.11.013}}.
\bibitem[{Raut et~al.(2022)Raut, Das, Gochhayat, and Das}]{Raut2022}
\bibinfo{author}{P.~Raut}, \bibinfo{author}{J.~Das},
  \bibinfo{author}{J.~Gochhayat}, \bibinfo{author}{K.~Das},
\newblock \bibinfo{title}{Influence of workforce agility on crisis management:
  Role of job characteristics and higher administrative support in public
  administration},
\newblock \bibinfo{journal}{Materials Today: Proceedings International
  Conference on Recent Advances in Mechanical Engineering and Nanomaterials}
  \bibinfo{volume}{61} (\bibinfo{year}{2022}) \bibinfo{pages}{647--652}.
  \bibinfo{note}{\url{https://doi.org/10.1016/j.matpr.2021.08.121}}.
\bibitem[{Castillo et~al.(2024)Castillo, Louveton, and
  Parenthoen}]{Castillo2024}
\bibinfo{author}{J.~Castillo}, \bibinfo{author}{N.~Louveton},
  \bibinfo{author}{M.~Parenthoen},
\newblock \bibinfo{title}{Immersive training in cyber crisis management: The
  case of the resistecc project},
\newblock in: \bibinfo{booktitle}{IHM’24 Adjunct: Adjunct Proceedings of the
  35th Conference on l’Interaction Humain-Machine},
  \bibinfo{organization}{AFIHM}, \bibinfo{address}{Paris, France},
  \bibinfo{year}{2024}.
\bibitem[{AFNOR(2022)}]{AFNOR2022}
\bibinfo{author}{AFNOR}, \bibinfo{title}{Cyber-résilience. Reconstruction du
  SI et continuité d'activité métiers en cas de cyberattaque paralysante},
  \bibinfo{type}{Technical Report} \bibinfo{number}{SPEC 2208}, AFNOR,
  \bibinfo{year}{2022}.
\bibitem[{Angafor et~al.(2020)Angafor, Yevseyeva, and He}]{Angafor2020}
\bibinfo{author}{G.~Angafor}, \bibinfo{author}{I.~Yevseyeva},
  \bibinfo{author}{Y.~He},
\newblock \bibinfo{title}{Game-based learning: a review of tabletop exercises
  for cybersecurity incident response training},
\newblock \bibinfo{journal}{Security and Privacy} \bibinfo{volume}{3}
  (\bibinfo{year}{2020}) \bibinfo{pages}{e126}.
  \bibinfo{note}{\url{https://doi.org/10.1002/spy2.126}}.
\bibitem[{Yamin et~al.(2020)Yamin, Katt, and Gkioulos}]{Yamin2020}
\bibinfo{author}{M.~Yamin}, \bibinfo{author}{B.~Katt},
  \bibinfo{author}{V.~Gkioulos},
\newblock \bibinfo{title}{Cyber ranges and security testbeds: scenarios,
  functions, tools and architecture},
\newblock \bibinfo{journal}{Comput Secur} \bibinfo{volume}{8}
  (\bibinfo{year}{2020}) \bibinfo{pages}{1--25}.
  \bibinfo{note}{\url{https://doi.org/10.1016/j.cose.2019.101636}}.
\bibitem[{Debatty and Mees(2019)}]{Debatty2019}
\bibinfo{author}{T.~Debatty}, \bibinfo{author}{W.~Mees},
\newblock \bibinfo{title}{Building a cyber range for training cyber defense
  situation wareness},
\newblock in: \bibinfo{booktitle}{Proc. International Conference on Military
  Communications and Information Systems (ICMCIS)}, \bibinfo{publisher}{IEEE},
  \bibinfo{year}{2019}, p. \bibinfo{pages}{1–6}.
\bibitem[{Smeets(2022)}]{Smeets2022}
\bibinfo{author}{M.~Smeets},
\newblock \bibinfo{title}{The role of military cyber exercises: A case study of
  locked shields},
\newblock in: \bibinfo{booktitle}{2022 14th International Conference on Cyber
  Conflict: Keep Moving! (CyCon)}, volume \bibinfo{volume}{700},
  \bibinfo{year}{2022}, pp. \bibinfo{pages}{9--25}.
  \bibinfo{note}{\url{https://doi.org/10.23919/CyCon55549.2022.9811018}}.
\bibitem[{Castillo et~al.(2025)Castillo, Louveton, and
  Parenthoen}]{Castillo2025}
\bibinfo{author}{J.~Castillo}, \bibinfo{author}{N.~Louveton},
  \bibinfo{author}{M.~Parenthoen},
\newblock \bibinfo{title}{{Psychological and Behavioral Aspects and System
  Dynamics: Insights from Exercises Using a Cyber Range}},
\newblock in: \bibinfo{booktitle}{{Lecture Notes in Computer Science.}}, volume
  \bibinfo{volume}{15999} of \textit{\bibinfo{series}{Availability, Reliability
  and Security (ARES)}}, \bibinfo{organization}{{Ghent University}},
  \bibinfo{address}{Belgium}, \bibinfo{year}{2025}, pp.
  \bibinfo{pages}{59--76}.
  \bibinfo{note}{\url{https://doi.org/10.1007/978-3-032-00644-8_4}, SharedIt
  \url{https://rdcu.be/eMiu7}}.
\bibitem[{Castillo et~al.(2026)Castillo, Parenthoen, and
  Louveton}]{Castillo2026}
\bibinfo{author}{J.~Castillo}, \bibinfo{author}{M.~Parenthoen},
  \bibinfo{author}{N.~Louveton},
\newblock \bibinfo{title}{A theoretical model for enhancing cyber crisis
  training in cyber ranges: bridging cognitive load, situational awareness, and
  immersion},
\newblock \bibinfo{journal}{Cognition Technology and Work} \bibinfo{volume}{28}
  (\bibinfo{year}{2026}) \bibinfo{pages}{437–453}.
  \bibinfo{note}{\url{https://doi.org/10.1007/s10111-025-00844-3}}.
\bibitem[{P\"{a}ij\"{a}nen et~al.(2021)P\"{a}ij\"{a}nen, Saharinen, Salonen,
  Sipola, Vykopal, and Kokkonen}]{Paijanen2021}
\bibinfo{author}{J.~P\"{a}ij\"{a}nen}, \bibinfo{author}{K.~Saharinen},
  \bibinfo{author}{J.~Salonen}, \bibinfo{author}{T.~Sipola},
  \bibinfo{author}{J.~Vykopal}, \bibinfo{author}{T.~Kokkonen},
\newblock \bibinfo{title}{Cyber range: preparing for crisis or something just
  for technical people?},
\newblock in: \bibinfo{editor}{T.~Eze}, \bibinfo{editor}{L.~Speakman},
  \bibinfo{editor}{C.~Onwubiko} (Eds.), \bibinfo{booktitle}{Proceedings of the
  20th European conference on cyber warfare and security (ECCWS)},
  \bibinfo{year}{2021}, p.~\bibinfo{pages}{13}.
\bibitem[{{NIST Cyber Range Project Team} and {NICE Community Coordinating
  Council, Cybersecurity Skills Competitions Community of
  Interest}(2023)}]{NIST2023}
\bibinfo{author}{{NIST Cyber Range Project Team}}, \bibinfo{author}{{NICE
  Community Coordinating Council, Cybersecurity Skills Competitions Community
  of Interest}}, \bibinfo{title}{The cyber range: A guide},
  \bibinfo{type}{Technical Report}, National Institute of Standards and
  Technology, \bibinfo{year}{2023}.
\bibitem[{{ECSO Working Group 5 – Skills {\&} Human Factors
  –}(2024)}]{ECSO2024}
\bibinfo{author}{{ECSO Working Group 5 – Skills {\&} Human Factors –}},
  \bibinfo{title}{White Paper: Cyber Exercise Scenario Development},
  \bibinfo{type}{Technical Report}, ECSO, \bibinfo{year}{2024}.
  \bibinfo{note}{\url{https://ecs-org.eu/?publications=white-paper-cyber-exercise-scenario-development}}.
\bibitem[{Katsantonis et~al.(2023)Katsantonis, Manikas, and
  Mavridis}]{Katsantonis2023}
\bibinfo{author}{M.~Katsantonis}, \bibinfo{author}{A.~Manikas},
  \bibinfo{author}{I.~Mavridis},
\newblock \bibinfo{title}{Cyber range design framework for cyber security
  education and training},
\newblock \bibinfo{journal}{Int. J. Inf. Secur.} \bibinfo{volume}{22}
  (\bibinfo{year}{2023}) \bibinfo{pages}{1005--1027}.
  \bibinfo{note}{\url{https://doi.org/10.1007/s10207-023-00680-4}}.
\bibitem[{Petersen et~al.(2020)Petersen, Santos, Smith, Wetzel, and
  Witte}]{Petersen2020}
\bibinfo{author}{R.~Petersen}, \bibinfo{author}{D.~Santos},
  \bibinfo{author}{M.~Smith}, \bibinfo{author}{K.~Wetzel},
  \bibinfo{author}{G.~Witte}, \bibinfo{title}{Workforce framework for
  cybersecurity (NICE Framework)}, \bibinfo{type}{Technical Report}
  \bibinfo{number}{800-181 r1}, National Institute of Standards and Technology,
  \bibinfo{year}{2020}.
\bibitem[{Raymondo and Loscacio(2024)}]{NIST2024}
\bibinfo{author}{G.~Raymondo}, \bibinfo{author}{L.~Loscacio},
  \bibinfo{title}{NIST cybersecurity framework 2.0.}, \bibinfo{type}{Technical
  Report}, National Institute of Standards and Technology,
  \bibinfo{year}{2024}.
\bibitem[{Zacharis et~al.(2026)Zacharis, Sarri, Heurck, Fanourakis, Fernández,
  Christoforatos, and Arcus}]{ENISA2026}
\bibinfo{author}{A.~Zacharis}, \bibinfo{author}{A.~Sarri},
  \bibinfo{author}{C.~V. Heurck}, \bibinfo{author}{F.~Fanourakis},
  \bibinfo{author}{G.~Fernández}, \bibinfo{author}{N.~Christoforatos},
  \bibinfo{author}{R.~Arcus}, \bibinfo{title}{The ENISA Cybersecurity Exercise
  Methodology. End-to-end guide on how to plan, run and evaluate an exercise},
  \bibinfo{type}{Technical Report} \bibinfo{number}{TP-01-25-028-EN-N}, ENISA,
  \bibinfo{year}{2026}. \bibinfo{note}{DOI 0.2824/4949728}.
\bibitem[{Castillo et~al.(2026)Castillo, Louveton, and
  Parentho\"{e}n}]{Castillo2026b}
\bibinfo{author}{J.~M. Castillo}, \bibinfo{author}{N.~Louveton},
  \bibinfo{author}{M.~Parentho\"{e}n},
\newblock \bibinfo{title}{Ergonomie cognitive et formation immersive : Aspects
  de la charge cognitive, de la conscience de la situation et de la présence
  dans la gestion des crises d'origine cyber},
\newblock in: \bibinfo{booktitle}{Proceedings du 13è colloque d'ergonomie
  cognitive EPIQUE}, \bibinfo{publisher}{Arpege Science Publishing},
  \bibinfo{address}{Lorient, France}, \bibinfo{year}{2026}.
\bibitem[{Zelter(1992)}]{Zelter92}
\bibinfo{author}{D.~Zelter},
\newblock \bibinfo{title}{Autonomy, interaction and presence},
\newblock \bibinfo{journal}{Presence} \bibinfo{volume}{1}
  (\bibinfo{year}{1992}) \bibinfo{pages}{127--132}.

\end{thebibliography}

\end{document}